\documentclass{cupconf}
\usepackage{natbib}
\usepackage{epsfig}
\usepackage{deluxetable}

  \checkfont{eurm10}
  \iffontfound
    \IfFileExists{upmath.sty}
      {\typeout{^^JFound AMS Euler Roman fonts on the system,
                   using the 'upmath' package.^^J}%
       \usepackage{upmath}}
      {\typeout{^^JFound AMS Euler Roman fonts on the system, but you
                   dont seem to have the}%
       \typeout{'upmath' package installed. cupconf.cls can take advantage
                 of these fonts,^^Jif you use 'upmath' package.^^J}%
      }
  \else
  \fi

  \checkfont{msam10}
  \iffontfound
    \IfFileExists{amssymb.sty}
      {\typeout{^^JFound AMS Symbol fonts on the system, using the
                'amssymb' package.^^J}%
       \usepackage{amssymb}%
         \let\leq=\leqslant
         
      }{}
  \fi

  \IfFileExists{amsbsy.sty}
    {\typeout{^^JFound the 'amsbsy' package on the system, using it.^^J}%
     \usepackage{amsbsy}}
    {}

\newcommand\eg{e.g.\ }

\title[Gravitational waves]{Gravitational waves from black-hole mergers}

\author[J. G. Baker {\it et al.\/}]%
{J\ls O\ls H\ls N\ns G.\ns B\ls A\ls K\ls E\ls R\ls$^1$,
W\ls I\ls L\ls L\ls I\ls A\ls M\ns D.\ns B\ls O\ls G\ls G\ls S\ls$^2$,
J\ls O\ls A\ls N\ns M.\ns C\ls E\ls N\ls T\ls R\ls E\ls L\ls L\ls A\ls$^1$,
B\ls E\ls R\ls N\ls A\ls R\ls D\ns J.\ns K\ls E\ls L\ls L\ls Y\ls$^1$,
S\ls E\ls A\ls N\ns T.\ns M\ls c\ls W\ls I\ls L\ls L\ls I\ls A\ls M\ls S\ls$^2$,
\and J\ls A\ls M\ls E\ls S\ns R.\ns v\ls a\ls n\ns M\ls E\ls T\ls E\ls R$^3$}

\affiliation{$^1$Gravitational Astrophysics Laboratory, NASA Goddard Space Flight Center, 8800 Greenbelt Rd., Greenbelt, MD 20771, USA\\[\affilskip]
$^2$University of Maryland, Department of Physics, 
College Park, MD 20742, USA\\[\affilskip]
$^3$Center for Space Science \& Technology, 
University of Maryland Baltimore
County, Physics Department, 1000 Hilltop Circle, Baltimore, MD 21250, USA}

\begin{document}

\maketitle

\begin{abstract}
  Coalescing black-hole binaries are expected to be the strongest
  sources of gravitational waves for ground-based interferometers as
  well as the space-based interferometer LISA. Recent progress in
  numerical relativity now makes it possible to calculate the
  waveforms from the strong-field dynamical merger and is
  revolutionizing our understanding of these systems.  We review these
  dramatic developments, emphasizing applications to issues in
  gravitational wave observations.  These new capabilities also make
  possible accurate calculations of the recoil or kick imparted to the
  final remnant black hole when the merging components have unequal
  masses, or unequal or unaligned spins.  We highlight recent work in
  this area, focusing on results of interest to astrophysics.
\end{abstract}

\firstsection 

\section{Introduction}

Gravitational wave astronomy will open a new observational window on
the universe.  Since large masses concentrated in small volumes and
moving at high velocities generate the strongest and therefore most
readily detectable waves, the final coalescence of black-hole binaries
is expected to be one of the strongest sources. The opening of the
full electromagnetic spectrum to astronomical observation during the
last century greatly expanded our understanding of the cosmos.  In
this new century, observations across the gravitational wave spectrum
will provide a wealth of new knowledge, including accurate
measurements of binary black-hole masses and spins.

The high frequency part of the gravitational wave spectrum, $\sim
10{\rm Hz} \lesssim f \lesssim 10^3{\rm Hz}$, is being opened today
through the pioneering efforts of first-generation ground-based
interferometers such as LIGO, currently operating at design
sensitivity.  Such instruments can detect gravitational waves from
coalescing stellar-mass ($M \lesssim 10^{2}M_{\odot}$) and
intermediate-mass ($10^{2}M_{\odot} \lesssim M \lesssim
10^{3}M_{\odot}$) black-hole binaries.  While detections from this
first generation of detectors are likely to be rare, the advanced LIGO
(adLIGO) upgrade may detect the coalescence of several stellar-mass
and tens of intermediate-mass black-hole binaries per year. Other
high-frequency sources include binary neutron star coalescences,
supernovae, and rotating neutron stars.

The low frequency gravitational wave window, $3\times 10^{-5}{\rm Hz}
\lesssim f \lesssim 1{\rm Hz}$, is especially rich in astrophysical
sources and will be opened by the space-based LISA detector, currently
in the formulation stage.  LISA will be sensitive to the coalescence
of massive black-hole binaries with total masses in the range
$10^4M_{\odot} \lesssim M \lesssim 10^7M_{\odot}$ to large redshifts
$z \gtrsim 10$ at relatively high signal-to-noise ratios (SNRs), and
may detect 10 or more such events per year.  Using such observations,
the black-hole masses, spins and luminosity distances can be
determined to very good precision, with errors $< 1\%$ in some cases
\citep{Lang:2006bz}. In addition, LISA will detect gravitational waves
from the inspiral of compact stars into central massive black holes
out to $z \sim 1$, as well as tens of thousands of compact binaries in
the Galaxy.

The actual merger of two comparable-mass black holes that plunge
together and form a common event horizon takes place in the
strong-field dynamical regime of general relativity.  For many years,
we were unable to calculate the expected waveforms from these very
energetic events due to severe problems with the large scale computer
codes needed to simulate the mergers. Recently, however, a series of
stunning breakthroughs has occurred in numerical relativity, resulting
in stable, robust, and accurate simulations of black-hole mergers as
well as applications to astrophysics.  In Sec.~\ref{sec:calc} we
review these developments and present examples of the resulting
gravitational waveforms.  Applications of these signals to issues in
gravitational wave observations are discussed in Sec.~\ref{sec:obs}.
When the merging black holes have unequal masses, or unequal or
unaligned spins, the final remnant black hole suffers a recoil; recent
progress in calculating these ``kicks'' and their applications to
astrophysics are presented in Sec.~\ref{sec:kicks}.  We conclude with
a summary in Sec.~\ref{sec:outlook}.

\section{Calculating black-hole binary coalescence}\label{sec:calc}

The final coalescence of a black-hole binary is driven by
gravitational wave emission, and proceeds in three stages: an
adiabatic inspiral, a dynamical merger, and a final ringdown
\citep{Flanagan97a}.  During the inspiral, the black holes are
well-separated and can be approximated as point particles.  The black
holes spiral together on quasicircular trajectories, and the resulting
gravitational waveforms are {\em chirps}, i.e.\ sinusoids that
increase in both frequency and amplitude as the black holes get closer
together. The inspiral can be treated analytically using the
post-Newtonian (PN) approach, which is an expansion in $v/c$, where
$v$ is the characteristic orbital velocity (see \cite{Blanchet06} for
a review of PN results).  The inspiral is followed by a dynamical
merger in which the black holes plunge together to form a highly
distorted single black hole, producing a powerful burst of
gravitational radiation.  Since the merger stage occurs in the regime
of very strong gravity, a full understanding of this process requires
numerical relativity simulations of the Einstein equations.  After
merger, the remnant black hole then settles down, evolving towards a
quiescent Kerr state by shedding its nonaxisymmetric modes as
gravitational waves.  The late part of this ringdown stage can be
treated analytically using black-hole perturbation theory, and the
resulting gravitational waveforms are superpositions of exponentially
damped sinusoids of constant frequency \citep{Leaver85,Echeverria89}.

In numerical relativity, the full set of Einstein's equations are
solved on a computer in the dynamical, nonlinear regime.  This is
typically accomplished by slicing 4-D spacetime into a stack of 3-D
spacelike hypersufaces, each labeled by time $t$
\citep{Arnowitt62,Misner73}.  The Einstein equations split into two
sets.  The constraints give a set of relationships that must hold on
each slice and in particular constrain the initial data for a
black-hole binary simulation.  This data is then propagated forward in
time using the evolution equations.  Four freely-specifiable
coordinate, or gauge, conditions give the development of the spatial
and temporal coordinates during the evolution.

Simulating the merger of a black-hole binary using numerical
relativity has proved to be very challenging.  The first attempt to
evolve a head-on collision in 2-D axisymmetry dates back to 1964
\citep{Hahn64}.  In the mid-1970s, the head-on collision of two equal
mass, nonspinning black holes was first simulated successfully, along
with the extraction of some information about the gravitational
radiation \citep{Smarr76,Smarr77,Smarr79}. In the 1990s, fully 3-D
numerical relativity codes were developed and used to evolve grazing
collisions of black holes \citep{Bruegmann97,Brandt00,Alcubierre00b}.
However, the codes were plagued by a host of instabilities that caused
them to crash before any significant portion of a black-hole binary
orbit could be evolved.  For many years, progress was slow and
incremental.

Recently, a series of dramatic developments has led to major progress
in black-hole binary simulations across a broad front.  The first
complete orbit of a black-hole binary was achieved in 2004
\citep{Bruegmann:2003aw}.  This was followed by the first full
simulation of a black-hole binary through an orbit, plunge, merger and
ringdown in 2005 \citep{Pretorius:2005gq}.  In late 2005, the
development of new coordinate conditions produced a breakthrough in
the ability to carry out accurate and stable long-term evolutions of
black-hole binaries
\citep{Campanelli:2005dd,Baker:2005vv,vanMeter:2006vi}.  These novel
but simple ``moving puncture'' techniques proved highly effective.
They were quickly adopted by a broad segment of the numerical
relativity community, leading to stunning advances in black-hole
binary modeling, starting with evolutions of equal mass, nonspinning
black holes and moving quickly to include unequal masses and spins;
(see, \eg, \cite{Campanelli:2006gf,Baker:2006yw,Campanelli:2006uy,
  Gonzalez:2006md,Baker:2006kr,Herrmann:2006ks,Herrmann:2007ac,
  Campanelli:2007ew,Campanelli:2007cg,Koppitz:2007ev,Gonzalez:2007hi,
  Baker:2007gi,Tichy:2007hk}).

The most rapid advances in modeling black-hole binary coalescences
cover the previously least understood part of the gravitational
waveform, i.e.\ the final few cycles of radiation generated from near
the ``innermost stable circular orbit'' (ISCO) and afterward, which we
call the ``merger-ringdown''.  There is already considerable progress
toward a full understanding of this important ``burst'' portion of the
waveform, through which the frequency sweeps by a factor of $\sim 3$
up to ringdown and during which the gravitational wave luminosity is
$\sim 10^{23}L_{\odot}$, more than the luminosity of the combined
starlight in the visible universe.

\begin{figure}
\includegraphics[width=13.8cm,clip=true,angle=0]{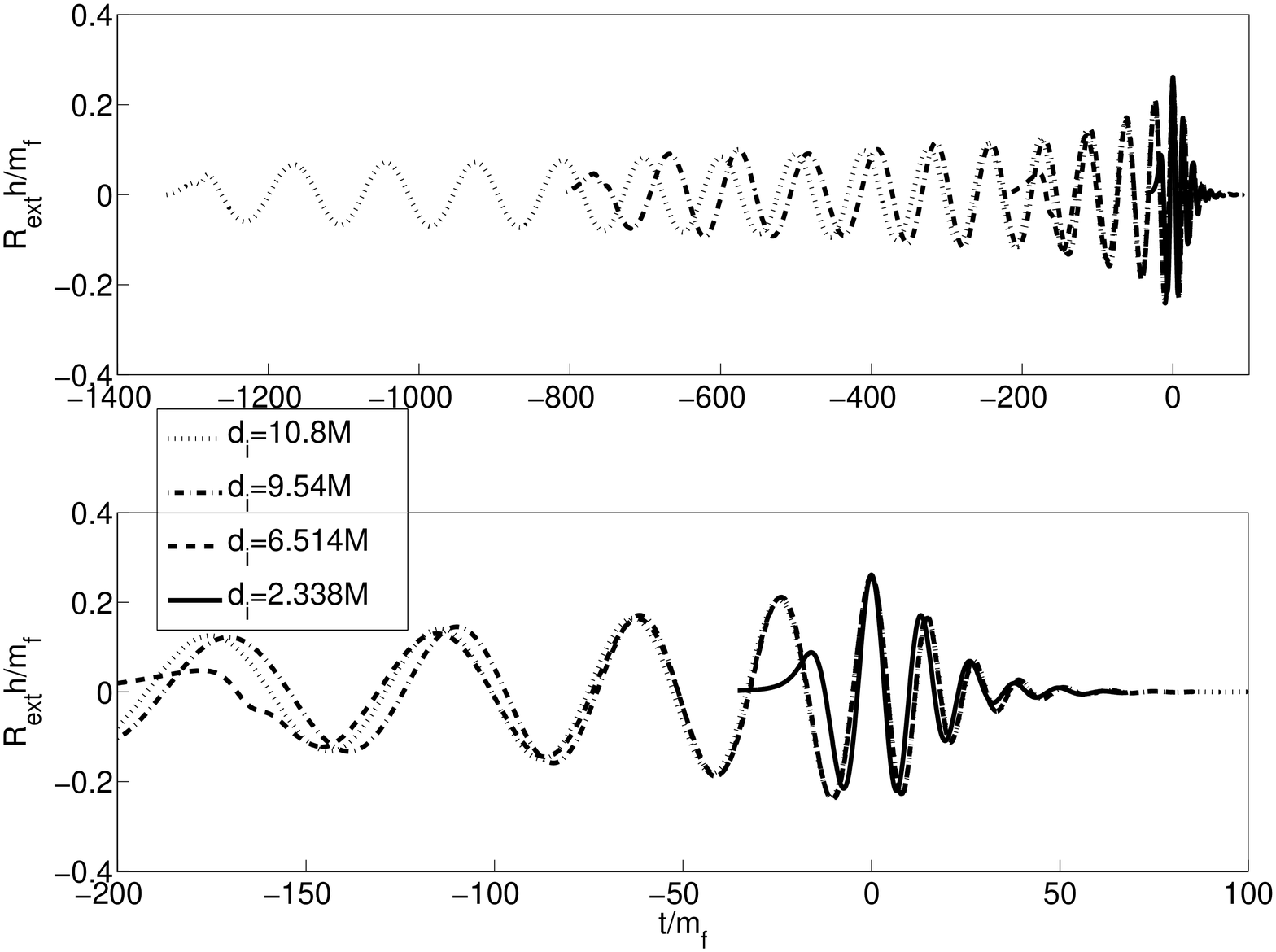}
\caption{Gravitational waveforms are shown for four simulations of
  equal mass nonspinning black-hole binaries with successively wider
  initial separations.  Here, $d_i$ is the initial coordinate
  separation of the black-hole centers (``punctures''). All waveforms
  have been aligned so that the moment of maximum radiation amplitude
  occurs at $t = 0$ (phase is also set to zero at $t = 0$).  The upper
  panel shows the full extent of all the runs, while the lower panel
  shows the final merger-ringdown burst.  }
\label{fig:waves}
\end{figure}

A particularly significant development was the demonstration of
initial-data-indepen\-dence of merger-ringdown waveforms for
equal-mass, nonspinning black holes \citep{Baker:2006yw}, as
summarized in Fig.~\ref{fig:waves}.
Results from four runs with successively larger initial separations
are shown; the waveforms have been aligned so that the moment of peak
radiation amplitude in each simulation occurs at time $t=0$.  Here we
show the gravitational wave strain from the dominant ($l=2$, $m=2$)
mode; this represents an observation made on the equatorial plane of
the system, where only a single polarization component contributes to
the measured strain. The upper panel of Fig.~\ref{fig:waves} shows the
full simulation waveforms, while the lower panel focuses on the final
merger-ringdown burst.  Note that here and elsewhere in this paper, we
use \emph{geometrical units}, $G = c = 1$, to measure time, distance
and mass in the same units.  In particular, one solar mass $M_{\odot}$
is equivalent to $\sim 5 \times 10^{-6}{\rm sec}$, or $\sim 1.5 {\rm
  km}$.\footnote{Since the simulation results scale with the masses of
  the black holes, they are equally applicable to LISA and
  ground-based detectors.}  In Fig.~\ref{fig:waves}, our timescale is
the final mass $m_f$, of the post-merger hole; this will be less than
the initial total binary mass $M$ because of gravitational radiation.
In the shortest run (solid line), the black holes are placed on
initial orbits close to the ISCO and undergo a brief plunge followed
by a merger and ringdown \citep{Baker:2005vv}. At the next-largest
initial separation (dashed line), the black holes complete $\sim 1.8$
orbits before merging \citep{Baker:2006yw}.  The waveforms from the
two runs with successively larger initial separations (dot-dashed and
dotted lines, respectively) then lock on to the merger-ringdown part
of this shorter (dashed) run.  In fact, the waveforms from the three
longer simulations show very strong agreement for $t \gtrsim
-50\,m_f$, with differences among these waveforms $\lesssim 1\%$ in
this regime.

Today, there is consensus that the merger of equal-mass, nonspinning
black holes produces a final Kerr black hole with spin $a/m_f \sim
0.7$, and that the amount of energy radiated in the form of
gravitational waves, starting with the final few orbits and proceeding
through the plunge, merger and ringdown, is $E_{\rm rad} = M - m_f
\sim 0.04M$
\citep{Pretorius:2005gq,Campanelli:2005dd,Baker:2005vv,Campanelli:2006gf,Baker:2006yw};
see \cite{Centrella:2006it} for a review.  There is also agreement on
the overall simple shape of the waveforms shown in
Fig.~\ref{fig:waves}, and detailed comparison of results among
numerical relativity groups has already begun \citep{Baker:comp}.
Parameter-space exploration of the merger-ringdown burst, using
simulations with various mass ratios
\citep{Herrmann:2006ks,Baker:2006vn,Gonzalez:2006md} and spins
\citep{Campanelli:2006uy,Campanelli:2006fg}, is now underway.  The
next step is to push this frontier to increasingly complex mass-ratio
and spin-orientation combinations, and to establish initial-data
independence across these parameters.

Simulations starting in the late-inspiral regime and covering more
than a few orbits prior to merger are more challenging than shorter
merger-ringdown evolutions \citep{Baker:2006kr}.  Such long
simulations have stronger requirements for numerical stability and
place greater demands on computational resources.  In addition, better
accuracy is needed to control the accumulation of phase error; this in
turn constrains the numerical error that can be tolerated in the rate
of energy loss through gravitational radiation reaction, which governs
the inspiral.

So far, the longest simulations have been carried out for equal-mass,
nonspinning black-hole binaries, starting at relatively wide
separations.  The two longest runs shown in Fig.~\ref{fig:waves}
undergo $\sim 4.2$ \citep{Baker:2006yw} and $\sim 7$ orbits
\citep{Baker:2006kr}, respectively, prior to merger\footnote{For these
  quadrupolar waves, the gravitational wave and orbital frequencies
  are related by $f = 2 f_{\rm orb}$.}  and demonstrate progress in
long simulations (see also \cite{Husa:2007hp}).  We expect that
astrophysical black-hole binaries in the late inspiral regime will
follow nearly circular orbits, since any initial eccentricity would
have been radiated away early in their evolution. However, in the
first long black-hole binary simulations, the eccentricity of the
initial orbits was not very well controlled; in particular, the early
part of the waveform shown by the dot-dashed line shows evidence of
eccentricity which resulted from the initial data specification.  The
dotted curve shows more recent results, which start with very small
eccentricity $\epsilon < 0.01$ (see also
\cite{Pfeiffer:2007yz,Husa:2007rh}).  Such long waveforms make it
possible to compare the results of numerical relativity simulations of
black-hole binaries with post-Newtonian calculations in the
late-inspiral regime \citep{Buonanno:2006ui,
  Baker:2006ha,Hannam:2007ik}, demonstrating remarkable agreement.

\section{Observing Black-Hole Binary mergers}\label{sec:obs}

The final moments of a black-hole binary merger produce the most
intense radiation generated through the strongest gravitational
dynamics. The resulting waveforms are obviously an important part of
the observable gravitational wave signature for black-hole binary
events.  Before these recent advances in numerical relativity,
however, little was confidently known about these signals, so that
observational analyses of gravitational wave data could only rely on
generic ``unmodeled burst'' techniques \citep{Abbott:2007wu} rather
than more effective matched-filtering techniques, which require
detailed knowledge of the expected signals.  Consequently the quality
of scientific information which could be expected from merger
observations has, to date, necessarily been significantly discounted
in planning observational work.

Just how much more we can learn from observing the final stages of
black-hole binary mergers with full knowledge of the waveforms, and
how best to apply this knowledge in gravitational wave data analysis,
are still largely unanswered questions, which the gravitational wave
community is only beginning to address.  We can get some rough
impressions of the observational significance of black-hole binary
mergers from this new perspective by considering the SNR of the full
waveforms based on matched-filtering analysis of gravitational wave
data \citep{Flanagan97a}.
\begin{figure}
\begin{center}
\leavevmode
\includegraphics*[scale=.35, angle=0]{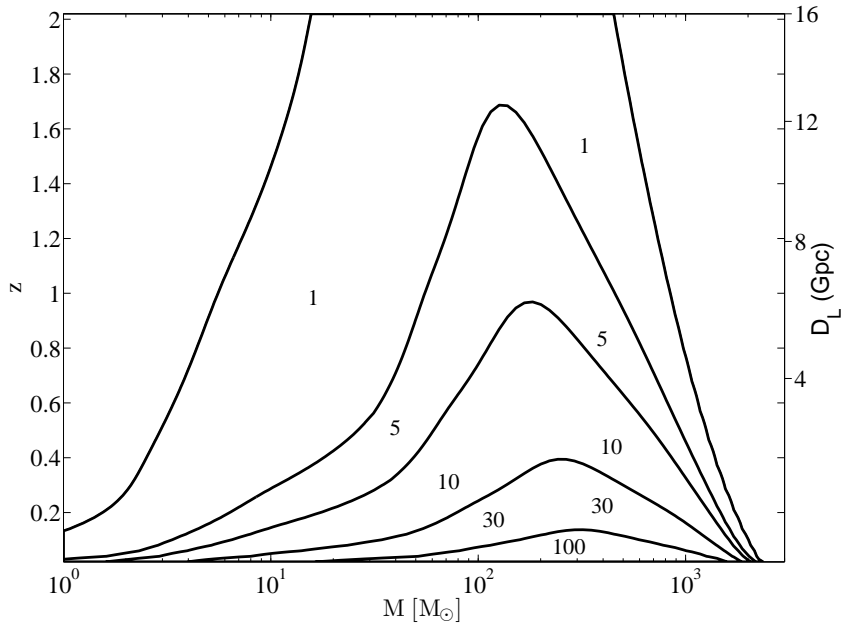}
\caption{Contours of SNR for observations of black-hole binaries with
  adLIGO, showing mass, redshift, and luminosity-distance dependence.
}
\label{fig:adLIGOSNRcontour}
\end{center}
\end{figure}
Estimating the full waveforms by stitching together the most accurate
PN model with a numerical simulation of the last several cycles, we
have calculated the SNR for equal-mass nonspinning binaries
\citep{Baker:2006kr}.  Fig.~\ref{fig:adLIGOSNRcontour} shows the
contours of SNR as functions of redshift $z$ and total binary mass $M$
for the ground-based adLIGO detector, which will be sensitive in the
frequency range $14Hz \lesssim f \lesssim 10^3$Hz. AdLIGO is a planned
upgrade of the initial LIGO detectors that will increase the
sensitivity by roughly an order of magnitude across the\ frequency
band. In addition, adLIGO can be tuned to optimize its sensitivity for
different sources.  To produce Fig.~\ref{fig:adLIGOSNRcontour} we used
the wide-band tuning typically associated with burst sources, due to
its greater sensitivity at higher frequencies, where the merger
portion from many sources is predicted to occur \citep{DHSCom}. This
gives an improved SNR for most black-hole masses compared to tunings
that were optimized for only the early inspiral portion of the
coalescence.
\begin{figure}
\begin{center}
\leavevmode
\includegraphics*[scale=.35, angle=0]{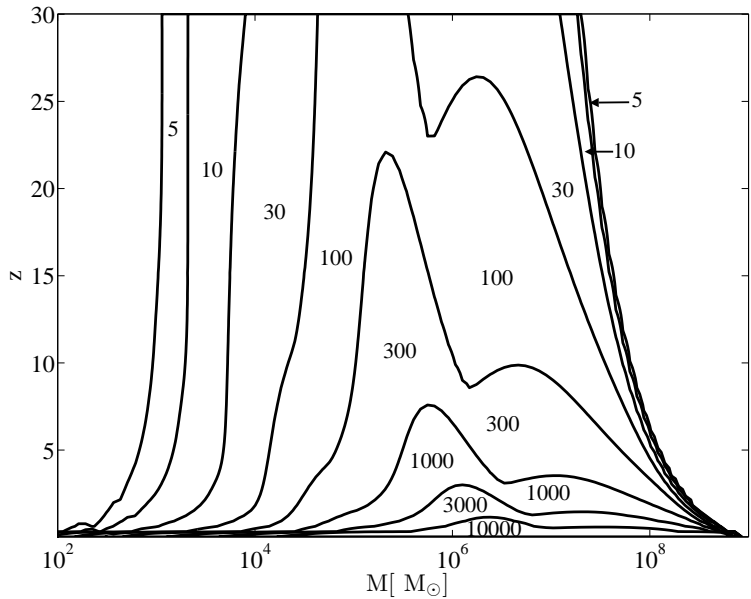}
\caption{Contours of LISA SNR with mass, redshift, and
  luminosity-distance dependence.  Note that black-hole binaries with
  masses $M > 10^7 M_{\odot}$ may not coalesce within a Hubble time
  \citep{Milosavljevic:2002bn}.}
\label{fig:LISASNRcontour}
\end{center}
\end{figure}
In Fig.~\ref{fig:LISASNRcontour} we plot contours of SNR for LISA
observations, showing that LISA can observe massive black-hole
binaries at high SNR throughout the observable universe.  LISA is most
sensitive to systems with masses in the range $10^5 M_{\odot} \leq M
\leq 10^7 M_{\odot}$, coinciding with the mass range in which models
of massive black-hole binaries predict coalescence within a Hubble
time \citep{Milosavljevic:2002bn} and event rates for LISA of at least
several per year \citep{Sesana:2004gf}.

\begin{figure}
\begin{center}
\leavevmode
\includegraphics*[scale=.35, angle=0]{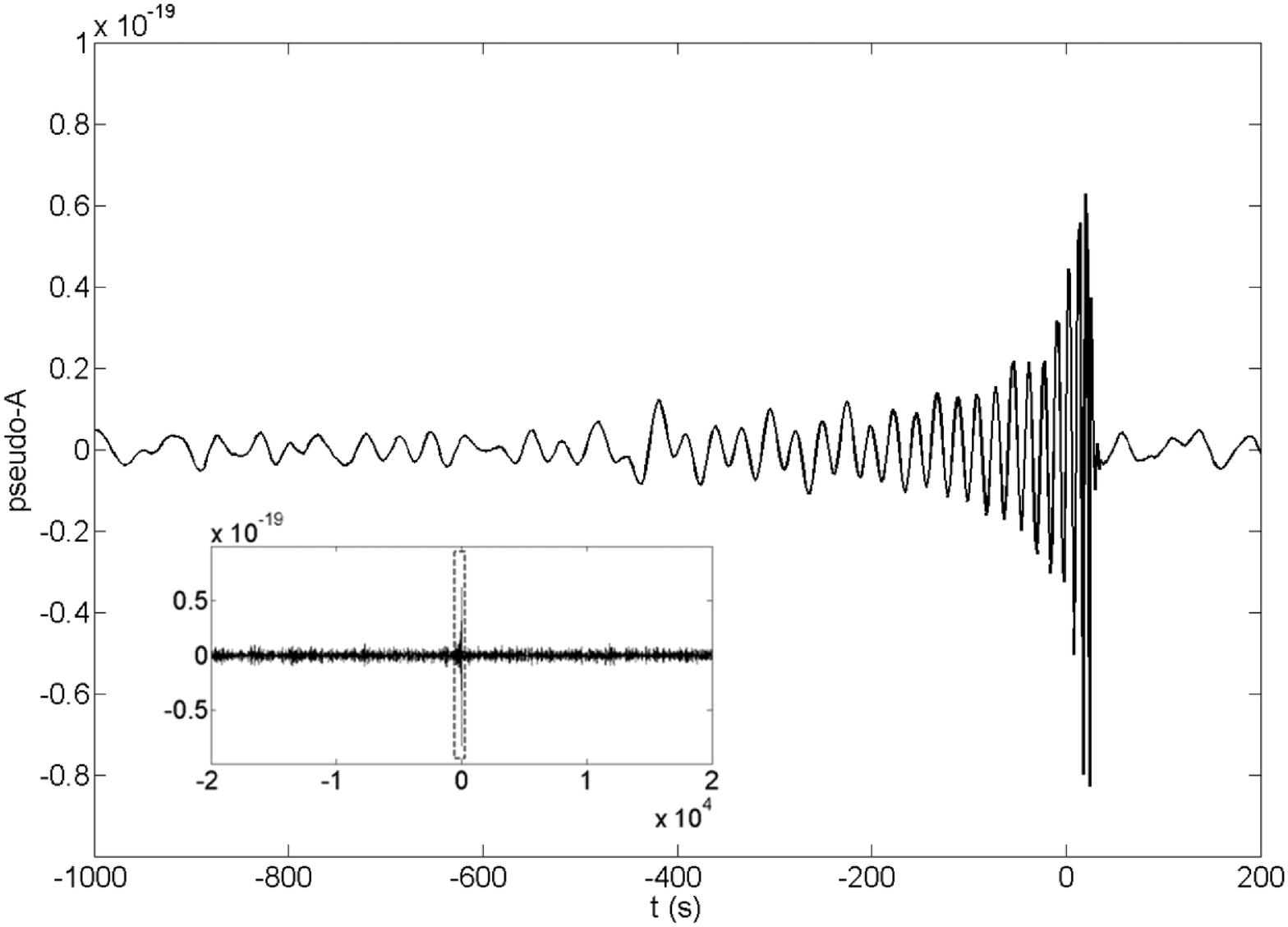}
\caption{Simulated LISA data stream showing LISA's response to a
  system of two equal-mass black holes. Specifically, the pseudo-A
  observable is shown for the case of two $4 \times 10^5 M_{\odot}$
  black holes at redshift $z = 10$, inclination $\iota = \pi/2$,
  polarization angle $\psi = 0$, ecliptic latitude $\beta = 0$, and
  ecliptic longitude $\lambda = \pi$.  The pseudo-A, E, and T
  observables are a set of linearly independent combinations of the
  Michelson unequal arm X, Y, and Z observables at TDI 1.5
  \citep{Shaddock:2003dj}, and are therefore useful in data analysis.
  The inset shows a time span of $4 \times 10^4$ seconds encompassing
  the time span of the main figure, the main figure being an
  enlargement of the dash-boxed region of the inset. }
\label{fig:LISAresponse}
\end{center}
\end{figure}

In Fig.~\ref{fig:LISAresponse} we plot a simulated LISA data stream
containing a black-hole binary signal as well as noise from the
interferometer and the white dwarf confusion noise caused from the
unresolvable number of background white dwarf binaries.  This is an
example of a typical, non-optimal case in order to demonstrate the
relative strength that these signals will generally have.  Here we
show two $4 \times 10^5 M_{\odot}$ black holes at redshift $z = 10$,
inclination $\iota = \pi/2$, polarization angle $\psi = 0$, ecliptic
latitude $\beta = 0$, and ecliptic longitude $\lambda = \pi$.
So far we have restricted our consideration to the example of
equal-mass nonspinning black-hole binary mergers.  More detailed
considerations of black-hole binary observations must take into
account the full parameter space of these systems, which, from the
point of view of general relativity, are described by seven
parameters, corresponding to the black-hole mass-ratio $q \equiv
m_1/m_2 \leq 1$ (recall that the total mass $M$ only scales the
waveforms), and the spin vector of each hole (plus two more parameters
if eccentric systems are considered).  Much progress has already been
made in conducting numerical simulations to flesh out the basic
features of this parameter space (see, \eg,
\cite{Campanelli:2006fg,Campanelli:2006uy,Herrmann:2006ks}).  First
impressions indicate that the waveforms in this parameter space are
qualitatively simple, with smooth parameter dependencies.  This
suggests that our knowledge of the mergers can ultimately be applied
in observations with the aid of smooth analytic waveform models,
joining PN treatments of the inspiral signals with analytic
approximations to the mergers.  Already such a model has been
proposed, which provides high overlaps with the numerical simulations
for mass-ratios as low as $q = 1/4$ \citep{Buonanno:2007pf}.

Applying the mass scalings from \cite{Flanagan97a}, we can estimate
the effect of the mass ratio $q$ on the computed SNRs; specifically,
SNR $\sim \eta^{1/2}$ for the inspiral, and SNR $\sim \eta$ for the
merger and ringdown, where $\eta \equiv m_1m_2/M^2 = q/(1+q)^2$ is the
\emph{symmetric mass ratio}.  For stellar-mass black-hole binaries,
the mass ratios are rather broadly distributed
\citep{Belczynski:2006zi}; the rates for such mergers may be low,
$\sim 2 yr^{-1}$ for adLIGO, depending on the evolution of the
original binary through the common envelope phase.  For
intermediate-mass black-hole binaries, mass ratios in the range $0.1
\lesssim q \lesssim 1$ are expected to be the most relevant, with
potential rates of $\sim 10$ per year \citep{Fregeau:2006yz}; we note,
however, that these rates are far more uncertain than those for
stellar-mass black-hole binaries.  Investigations of
gravitational-wave data analysis based on unequal-mass numerical
simulations are now underway: \cite{Pan:2007nw} have investigated how
numerical relativity waveforms may be used to select the best
post-Newtonian template banks for searches in the LIGO, adLIGO and
VIRGO data stream.

Astrophysical black-hole binaries are also expected to be spinning,
which can potentially raise the SNR, for example if there is a
spin-induced hangup that generates more gravitational wave cycles in
the merger \citep{Campanelli:2006uy}.  \cite{Vaishnav:2007nm} have
investigated how the addition of spins to equal-mass waveforms affects
search templates, necessitating the use of higher-order modes for high
matches in the LIGO band.

We also note that, with more accurate computational modeling,
simulations of the merger can be used together with gravitational wave
observations to probe gravity in the regime of strong fields.  In
particular, if the binary masses and spins can be obtained with good
accuracy from observations of the inspiral with LISA
\citep{Lang:2006bz}, the merger waveform can be calculated using
numerical relativity. With this, a comparison can be made between the
predictions of general relativity -- or any other theory of gravity
used in a numerical simulation -- with observations in the regime of
very strong gravity.

\section{Kicks from black-hole binary mergers}\label{sec:kicks}

Black-hole binaries that possess an asymmetry, such as having unequal
masses or possessing unequal or unaligned spins, will emit
gravitational radiation asymmetrically.  This asymmetric gravitational
radiation will impart a linear momentum recoil on the binary's center
of mass.  Previous attempts to calculate this recoil, or ``kick'',
analytically in the nonspinning case \citep{Fitchett_1983} and for
cases with spin \citep{Kidder95a} included very large uncertainties,
because the dominant part of the kick occurred in the last orbit
through the merger, which was very poorly modeled.  With the advent of
stable, accurate numerical relativity codes, the problem of accurately
calculating the kick became possible.  Several groups have tackled
this problem.  The problem of kicks from an unequal mass binary was
first investigated in \cite{Baker:2006vn}, where recoils between 86
and 115 km/s for a $q=2/3$ mass ratio were simulated.  In
\cite{Gonzalez:2006md}, several simulations were performed and a peak
recoil velocity for nonspinning binaries of 175 km/s was found for a
mass ratio of $q \approx 1/2.78$.

A flurry of activity began when several groups began simulating kicks
from spin asymmetry.  \cite{Herrmann:2007ac} demonstrated that kicks
from spin asymmetry can exceed unequal-mass kicks, and that for an
equal-mass configuration with one black hole spinning prograde and the
other spinning retrograde with equal magnitude $\hat{a}$ (a
dimensionless spin parameter, varying from 0 for a Schwarzschild hole
to 1 for a maximally spinning Kerr hole), the spin kick scales as $475
\,\hat{a}$ km/s.  However, it was soon found that the largest kicks
are obtained when the spins are antiparallel with each other but
perpendicular to the orbital angular momentum, a configuration
suggested by \cite{Campanelli:2007ew}. \cite{Gonzalez:2007hi}
were the first to perform this simulation, obtaining a kick of $2650$
km/s for $\hat{a} = 0.8$.  \cite{Campanelli:2007cg} showed that the
resulting kick depends sensitively on the angle between the black-hole
spin vectors and their velocities.  They found a simple cosine
dependence, indicating that \cite{Gonzalez:2007hi} somewhat
fortuitously happened upon what is nearly the maximum angle.
Furthermore, \cite{Campanelli:2007cg} projected that for maximal spin,
kicks as large as $4000$ km/s are possible.

Kicks of $2000$ km/s are sufficiently large to kick remnant black
holes out of even the largest giant elliptical galaxies.  Therefore,
simulation results this large had to be reconciled with the
observational reality that we see massive black holes at the centers
of all the galaxies that we have observed above a certain spheroidal
mass threshold.  \cite*{Schnittman:2007sn} used the EOB formalism to
perform a Monte Carlo simulation in order to try and predict how
typical these large kicks are.  For mass ratios between $q=1$ and
$q=1/10$ and spins of $\hat{a} = 0.9$ on both holes, they allowed
random orientations among the spin vectors and the orbital angular
momentum.  They found that 12\% of the remnants received a kick
exceeding $500$ km/s, and 2.7\% received a kick exceeding $1000$ km/s.

\cite{Schnittman:2007sn} assume no external influence which might tend
to align the spins with each other and with the orbital angular
momentum.  However, \cite*{Bogdanovic:2007} investigated the role of
the Bardeen-Petterson effect \citep{BardeenPetterson_1974} in
mitigating the gravitational recoil from spin asymmetry.  In the
absence of gas accretion, the assumption of random orientation in
\cite{Schnittman:2007sn} applies.  However, if a black-hole binary
accretes (1-10)\% of its initial mass from an accretion disk, the spin
of each hole will align with the angular momentum of the disk, which
in turn is aligned with the orbital angular momentum of the binary.
Therefore, if the majority of mergers occur in gas-rich environments,
then the giant kick results, although theoretically interesting, are
not astrophysically relevant.  There is good reason to expect
supermassive black holes to accrete substantial mass, particularly at
larger redshift.  In this scenario, the configuration of greatest
interest is spin orientations aligned with the orbital angular
momentum.

\begin{figure}
\begin{center}
\leavevmode
\includegraphics*[scale=0.5,angle=0]{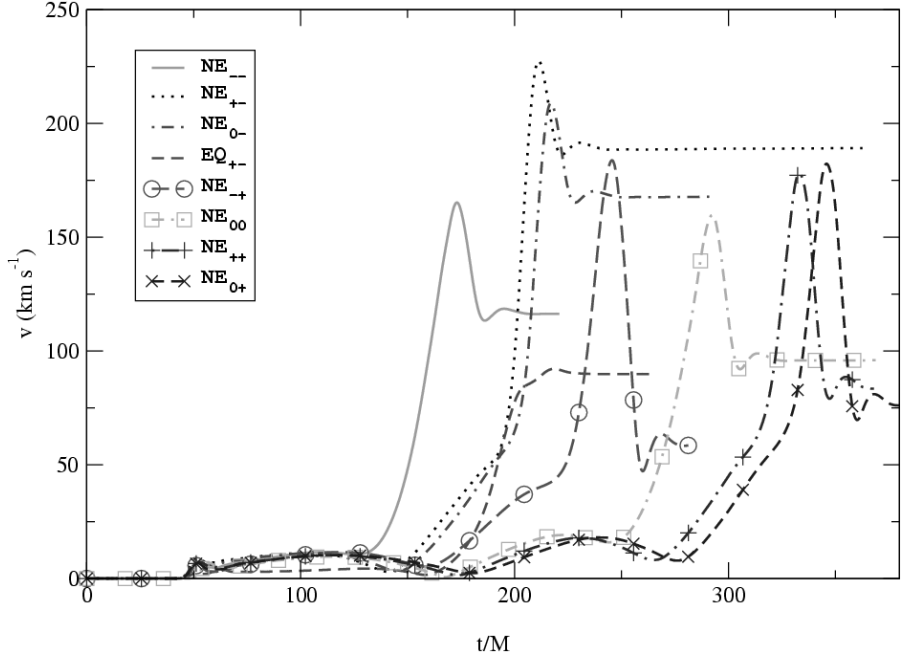}
\caption{Aggregated kicks from all of our runs.  The merger time for
  each binary matches the peak in its kick profile; the relative delay
  in merger times between data sets differing in initial spins is
  consistent with the results of \cite{Campanelli:2006uy}.  All
  configurations show a marked ``unkick'' after the peak, with the
  exception of the equal-mass case, ${\rm EQ}_{+-}$. [from
  \cite{Baker:2007gi} - reproduced with permission of the AAS]}
\label{fig:kicks}
\end{center}
\end{figure}

Fig. \ref{fig:kicks} and Table \ref{tab:data} present the results of
our investigation of this class of configurations, which were first
presented in \cite{Baker:2007gi}.  ``NE'' refers to unequal mass, with
the corresponding simulations having a mass ratio $q = 2/3$.  ``+''
means prograde with respect to the orbital angular momentum, and ``-''
means retrograde (``0'' means no spin).  Fig. \ref{fig:kicks} shows
the accumulated kick as a function of time for all of our runs.  Of
particular note is the absence of an ``unkick'' for the equal-mass
case, and the variable size of the unkick in the other cases.  Also,
we observe the expected accelerated merger in the ${\rm NE}_{--}$ and
${\rm NE}_{0-}$ cases due to spin-orbit attraction, and
correspondingly the delayed merger in the ${\rm NE}_{++}$ and ${\rm
  NE}_{0+}$ cases due to spin-orbit repulsion.  Using our data along
with the data from \cite{Herrmann:2007ac} and \cite{Koppitz:2007ev}
(which are also included in Table \ref{tab:data}), we are able to
construct an empirical kick formula, given by:
\begin{equation}
v = V_0[32\,q^2/(1+q)^5] \sqrt{(1-q)^2 + 2\,(1-q)\,K\,\cos\theta + K^2 },
\label{eqn:kick_formula}
\end{equation}
where $K = k\,(q\hat{a}_1-\hat{a}_2)$. The parameter $V_0$ gives the
overall scaling of the kick (note that the factor in brackets becomes
unity for $q=1$), while $k$ gives the relative scaling of the kick
contributions from spin and mass asymmetries.
\begin{center}
\begin{deluxetable}{lrrrrrr}
  \tablecolumns{7} \tablewidth{0pt}
  \tablecaption{Predicted versus computed kick speed.  Runs labeled
    ``S0.\#\#'' are taken from \cite{Herrmann:2007ac}, while runs
    labeled ``r\#'' are taken from \cite{Koppitz:2007ev}.  [from
    \cite{Baker:2007gi} - reproduced with permission of the
    AAS]\label{tab:results}}
  \tablehead{\colhead{Run} & \colhead{$q$}
    & \colhead{$\hat{a}_1$} & \colhead{$\hat{a}_2$} &
    \colhead{$v_{num}$} & \colhead{$v_{pred}$} &
    \colhead{$\frac{|\Delta v|}{v_{num}}$(\%)}} \startdata
  ${\rm NE}_{--}$ & 0.654 &-0.201 &-0.194 &116.3 &119.5 & 2.7\\
  ${\rm NE}_{-+}$ & 0.653 &-0.201 & 0.193 & 58.5 & 58.2 & 0.5\\
  ${\rm NE}_{0-}$ & 0.645 & 0.000 &-0.195 &167.7 &153.1 & 8.7\\
  ${\rm NE}_{00}$ & 0.677 & 0.000 & 0.000 & 95.8 & 98.6 & 2.9\\
  ${\rm NE}_{0+}$ & 0.645 & 0.000 & 0.194 & 76.9 & 71.7 & 6.8\\
  ${\rm NE}_{+-}$ & 0.655 & 0.201 &-0.194 &188.6 &181.9 & 3.6\\
  ${\rm NE}_{++}$ & 0.654 & 0.201 & 0.194 & 83.4 & 92.4 &10.8\\
  \medskip
  ${\rm EQ}_{+-}$ & 1.001 & 0.198 &-0.198 & 89.8 & 92.6 & 3.2\\
  S0.05 & 1.000 & 0.200 &-0.200 & 96.0 & 93.8 & 2.3\\
  S0.10 & 1.000 & 0.400 &-0.400 &190.0 &187.6 & 1.2\\
  S0.15 & 1.000 & 0.600 &-0.600 &285.0 &281.5 & 1.2\\
  \medskip
  S0.20 & 1.000 & 0.800 &-0.800 &392.0 &375.3 & 4.3\\
  r0 & 1.000 &-0.584 & 0.584 &260.0 &274.0 & 5.4\\
  r1 & 0.917 &-0.438 & 0.584 &220.0 &220.8 & 0.3\\
  r2 & 0.872 &-0.292 & 0.584 &190.0 &178.1 & 6.3\\
  r3 & 0.848 &-0.146 & 0.584 &140.0 &141.9 & 1.4\\
  r4 & 0.841 & 0.000 & 0.584 &105.0 &110.4 & 5.1\\
\enddata
\label{tab:data}
\end{deluxetable}
\end{center}

Performing a least-squares fit to the data yields $V_0 = 276$ km/s, $q
= 0.58$, and $k = 0.85$.  Our formula yields a maximum error of 10.8\%
for the cases investigated.  If, in fact, the majority of supermassive
black-hole mergers are gas-rich, then this formula may predict a
significant component of the kick for the majority of cases of
astrophysical interest.  \cite{Campanelli:2007cg} subsequently
suggested an extension of the formula presented in \cite{Baker:2007gi}
for kicks out of the orbital plane, which would be more applicable in
gas-poor cases where the Bardeen-Petterson effect doesn't occur, such
as gas-poor galaxies and the merger of stellar-mass black holes.  The
achievement of simulating the merger is therefore leading us to a
fairly comprehensive understanding of the recoil velocity parameter
space in a strikingly brief period of time.

\section{Summary}\label{sec:outlook}

Mergers of comparable-mass black-hole binaries produce intense bursts
of gravitational radiation and are very strong sources for both
ground-based interferometers and the space-based LISA. Recent progress
in numerical relativity now makes it possible to calculate the merger
waveforms accurately and robustly.  Today there is consensus that the
merger of equal-mass, nonspinning black holes produces a final Kerr
black hole with spin $a/m_f \sim 0.7$, and that the amount of energy
radiated in the form of gravitational waves, starting with the final
few orbits and proceeding through the plunge, merger and ringdown, is
$E_{\rm rad} = M - m_f \sim 0.04M$.  Simulations are now being carried
out for an increasing range of black-hole component masses and spins.
Computational methods continue to improve, resulting in longer
simulations that start in the late inspiral regime and allow
quantitative comparisons with analytic post-Newtonian methods.
Applications of the resulting waveforms to gravitational wave data
analysis have begun.  The recoil or kick velocity that results when
the black holes have unequal masses, or unequal or unaligned spins,
has been calculated for a variety of interesting cases, with key
applications to astrophysics.  Overall, the field of black-hole binary
mergers is experiencing a true ``golden age,'' with many new results
coming out across a broad front.  Stay tuned!

\begin{acknowledgments}
  It is a pleasure to thank our colleagues and collaborators
  Alessandra Buonanno, Scott Hughes, Cole Miller, Jeremy Schnittman
  and Tuck Stebbins for stimulating discussions.  This work was
  supported in part by NASA grant 06-BEFS06-19.  The simulations were
  carried out using Project Columbia at the NASA Advanced
  Supercomputing Division (Ames Research Center), and at the NASA
  Center for Computational Sciences (Goddard Space Flight Center).
  B.J.K. was supported by the NASA Postdoctoral Program at the Oak
  Ridge Associated Universities.  S.T.M. was supported in part by the
  Leon A. Herreid Graduate Fellowship.
\end{acknowledgments}

\end{document}